\documentclass[twocolumn,showpacs,prl,amsmath]{revtex4}
\usepackage[dvips]{graphicx}
\usepackage{psfig,epsfig,pstricks}
\usepackage{comment}
\usepackage{cases}
\usepackage{amsmath}
\usepackage{xspace}

\newcommand{\tm}{s}

\newcommand{\etal}{{\it et~al.}\xspace}
\newcommand{\ie}{{\it i.e.}\xspace}
\newcommand{\teff}{t_{\text{eff}}}
\newcommand{\ave}[1]{\left\langle#1 \right\rangle}
\newcommand{\FSS}{FSS\xspace}

\newcommand{\elabel}[1]{\label{eq:#1}}
\newcommand{\eref}[1]{(\ref{eq:#1})}
\newcommand{\Eref}[1]{Eq.~(\ref{eq:#1})}

\newcommand{\flabel}[1]{\label{fig:#1}}

\newcommand{\Fref}[1]{Fig.~\ref{fig:#1}}

\begin{document}
\title{Self-organized Criticality and Absorbing States: Lessons from the Ising
Model}
\date{\today}
\author{Gunnar Pruessner}
\email{gunnar.prussner@physics.org}
\homepage{http://www.ma.ic.ac.uk/~pruess}
\affiliation{Physics Department, Virginia Polytechnic Inst. \& State
Univ., Blacksburg, VA 24061-0435, USA}

\author{Ole Peters}
\email{ole.peters@ic.ac.uk}
\homepage{http://www.cmth.ph.ic.ac.uk/people/ole.peters}
\affiliation{Santa Fe Institute, 1399 Hyde Park Road, Santa Fe, NM 87501, USA\\
CNLS, Los Alamos National Laboratory, MS-B258, Los Alamos, NM 87545, USA}

\begin{abstract} 
We investigate a suggested path to self-organized
criticality. Originally, this path was devised to ``generate
criticality'' in systems displaying an absorbing-state phase
transition, but closer examination of the mechanism reveals that it
can be used for any continuous phase transition. We used the Ising
model as well as the Manna model to demonstrate how the finite-size
scaling exponents depend on the tuning of driving and dissipation
rates with system size. Our findings limit the explanatory power of
the mechanism to non-universal critical behavior.
\end{abstract}

\pacs{
05.65.+b, 
05.70.Jk, 
64.60.Ht 
}
\maketitle 
Self-organized criticality (SOC) refers to the spontaneous emergence of
critical behavior in slowly driven dissipative systems
\cite{BakTangWiesenfeld:1987,Jensen:1998}. 
Most models are defined on lattices with local particle numbers $z_i$
and thresholds $z_i^c$. They are driven discretely in time by
increasing $z_i$ at randomly chosen positions $i$ until such an
increase leads to $z_i>z_i^c$ somewhere in the system.  Particles then
topple to neighboring sites and can trigger avalanches of local
redistribution propagating through the entire lattice. Dissipation
typically takes place at the boundaries, where particles leave the
system. When an avalanche has finished the model is driven again 
\cite{Jensen:1998}. The resulting avalanche size distributions 
obey simple scaling.

In models displaying absorbing state (AS) phase transitions
\cite{Hinrichsen:2000} a tuning parameter, such as the overall
particle density, controls a transition between an inactive phase and
a phase where activity in the system continues indefinitely.

From the first introduction of SOC in 1987
\cite{BakTangWiesenfeld:1987}, it was believed that SOC models
manoeuver themselves to the critical density between similar inactive
and active phases.  Tang and Bak suggested in 1988 that the density of
``lattice sites on which $z>z_c$ [...] may be viewed as the order
parameter for this critical phenomenon'' \cite{Tang88b}. Such an
identification of the \emph{activity} with the order parameter implies
a link to absorbing state phase transitions.

This link was formalized and made explicit about 10 years later
\cite{VespignaniZapperi:1997,DickmanVespignaniZapperi:1998,DickmanETAL:2000,Dickman:2002}.
Dickman \etal \cite{DickmanVespignaniZapperi:1998} introduced periodic
boundaries to SOC systems, thereby turning them into AS models.
Measuring the exponents characterizing the spreading of perturbations
\cite{VespignaniETAL:2000,ChessaMarinariVespignani:1998} or the
roughness of the associated interface models
\cite{VespignaniETAL:2000,DickmanETAL:2001}, it has been observed that
at the critical density the closed-model behavior resembles that of
open SOC models
\cite{DickmanVespignaniZapperi:1998,ChristensenETAL:2004}.

The resulting interpretation of SOC is obvious
\cite{DickmanVespignaniZapperi:1998,DickmanETAL:2000,Dickman:2002}:
Activity eventually leads to dissipation at the boundaries, which in
turn reduces the particle density to below the critical value.
Driving takes place whenever quiescence has been reached. SOC models
therefore hover around the critical point, being pushed forth into the
active state by driving and pushed back into the quiescent state by
dissipation.

With this simple picture in mind one arrives at an equation of motion
for the particle density $\zeta$ in the system
\cite{VespignaniETAL:1998,VespignaniZapperi:1998}
\begin{equation}
\elabel{motion}
\frac{d}{d\tm} \zeta (\tm)=h-\rho_{a}(\tm)\epsilon \ , 
\end{equation}
where $\tm$ is the time, $h$ is the driving rate and $\epsilon$ is
called the (bulk) dissipation rate.
The activity
$\rho_{a}$ is the order parameter, defined as the density of active
sites, $z_i>z_i^c$, in the active phase. 
We will refer to this
interpretation of SOC as ``the AS approach''.

Clearly, the driving $h$ must be very slow compared to the dissipation
$\rho_a \epsilon$.  Otherwise particles would be added while the
system is active, leading to a fluctuating activity rather than
distinct avalanches. The proponents of the AS approach point out that
$h$, $\epsilon$ and $h/\epsilon$ have to be tuned to zero in order to
achieve the desired separation of timescales
\cite{VespignaniZapperi:1997}. 
While the definitions of SOC models typically restrict dissipation to
boundary sites and result in diverging avalanche sizes in the
thermodynamic limit, leading to appropriately vanishing $\epsilon(L)$
and $h(L)$, so far no statement has been made as to how the limiting
behaviour is approached. But this turns out to be the all-important
piece of information: The finite-size scaling (\FSS) behavior, the
only scaling available in SOC, depends entirely on the scaling of the
driving and dissipation rates with system size. Choosing $h(L)$ and
$\epsilon(L)$ freely, \emph{arbitrary} scaling behavior is produced.

In the following the relation between the scaling of $h$ and
$\epsilon$ and the resulting \FSS is analyzed, using the two
dimensional Ising model as an example. However, the analysis is
generally applicable and works equally well for standard SOC models
and their AS counterparts, which is confirmed by simulations of the
Manna model \cite{Manna:1991,DickmanETAL:2001}.

Translating \Eref{motion} into magnetic language, $\zeta$ corresponds
to the inverse temperature $\beta$ and the activity $\rho_{a}$ to the
modulus of the magnetization density $|m|$.  The parameters $h$ and
$\epsilon$ become cooling and heating rates, so that the temperature
$T$ is increased for large magnetizations and lowered otherwise,
\begin{equation}
\elabel{isingmotion}
\frac{d}{d\tm} \beta(\tm)=h-|m(\tm)|\epsilon \ .
\end{equation}
The resulting model is an Ising model where the temperature is
dynamically adapted according to the equation of motion
\eref{isingmotion}.
Therefore, the configurations are not sampled with Boltzmann-weight
and the resulting ``dynamical ensemble'' is not canonical.  
However,
by multiplying \eref{isingmotion} by a small pre-factor, corresponding
to rescaling the time, the distribution of temperatures can be made
arbitrarily narrow. 
For the sake of the following analysis, it is
assumed that this ``dynamical Ising model'' is well characterized by a
single effective (reduced) temperature, $\teff$.

For the \FSS analysis presented below we choose the approach of
$h,\epsilon \to 0$ to leading order as
\begin{equation}
h = h_0 L^{-\omega}\  \text{and}\ 
\epsilon = \epsilon_0 L^{-\kappa},
\elabel{ratescaling} 
\end{equation}
where $\omega, \kappa>0$. In the stationary state,
$\ave{\frac{d}{d\tm} \beta}=0$, \eref{isingmotion} yields
$\ave{|m|}=(h_0/\epsilon_0) L^{\kappa-\omega}$ with $\ave{}$ denoting
the average over the dynamical ensemble introduced above.  Clearly one
must choose $\omega > \kappa$.  To attain the prescribed
$\ave{|m|}(L)$ the system settles at the effective (reduced)
temperature $\ave{T-T_c}/T_c\equiv\teff(L)\propto L^{-1/\mu}$ to
leading order, see \Fref{cartoon}. Via $\teff(L)$ all
thermodynamic quantities depend only on $L$, which can be mistaken for
standard \FSS at temperature $T=T_c$. For the study of SOC models it
is vital to understand the difference
because SOC systems are always critical, wherefore \FSS is the only 
scaling available.

\begin{figure}
\includegraphics*[width=8cm,angle=0]{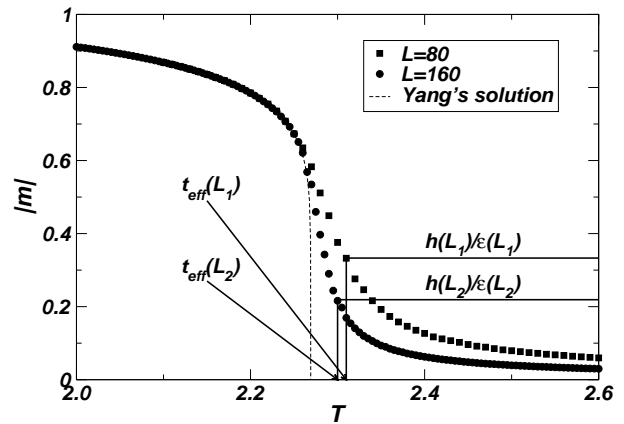}
\caption{Finite size behaviour for the magnetisation $|m|$ versus the
temperature $T$ in the regular, 2D-Ising model for $L_1=80$ (squares) and
$L_2=160$ (circles).  The dashed line shows Yang's solution \cite{Yang:1952}.
In the dynamical Ising model, the cooling and
heating rates, $h(L)$ and $\epsilon(L)$ prescribe the magnetisation
$|m(L)|=h(L)/\epsilon(L) \propto L^{\kappa-\omega}$, indicated by horizontal
lines. The system is forced to move to an effective temperature
$t_{\text{eff}}(L)$, indicated by arrows.}
\flabel{cartoon}
\end{figure}

Around the critical point of a continuous phase transition, the
singular part of the free energy leads to a simple scaling behavior of
the magnetization density \cite{PrivmanHohenbergAharony:1991},
\begin{equation}
\ave{|m|}=-k_hL^{-\beta/\nu}Y'(k_ttL^{1/\nu}),
\elabel{freeenergy}
\end{equation}
where $t$ is the reduced temperature, negative in the low temperature
phase (LTP) and positive in the high temperature phase (HTP).  $k_h$
and $k_t$ are metric factors,
and $Y'(x)$ is a universal scaling function, which becomes dependent on
the boundary conditions and the geometry of the system in the limit of
small arguments, case \eref{scaling_FSS} below.
There are 
three qualitatively different (asymptotic) regimes
\begin{subnumcases}{Y'(x) \to \elabel{scaling_fct}}
\propto |x|^\beta     & \text{for $x \to -\infty$} \elabel{scaling_LTP}\\
\text{const.}         & \text{for $x \to 0 $}      \elabel{scaling_FSS} \\
\propto x^{-\gamma/2} & \text{for $x \to \infty$}  \elabel{scaling_HTP}, 
\end{subnumcases}
where in the Ising model $\gamma=\nu d - 2 \beta$.

The first line describes the asymptotic behavior of the magnetization
in the LTP, the second line represents \FSS, and the third line
describes the HTP.

Setting $\ave{|m|}\propto L^{\kappa-\omega}$, the different regimes of
\eref{scaling_fct} are accessed by three qualitatively different
choices of $\kappa-\omega$, that is speeds at which $\ave{|m|}$
approaches zero:
\begin{itemize}
\item[1)] $\kappa-\omega>-\beta/\nu$ (``too slow''): In this case the
        magnetization approaches $0$ slower than in a standard
	Ising model kept at temperature $T=T_c$ as the system size
	increases, so that $Y'(k_{t}\teff(L) L^{1/\nu}) \propto
	L^{\kappa-\omega+\beta/\nu}$ is divergent in $L$. The only way
	to obtain a divergent $Y'(x)$ is via \eref{scaling_LTP}, which
	requires a negatively divergent argument $x \to - \infty$.
	The effective temperature is therefore negative and scales
	like $|\teff(L)|^\beta \propto L^{\kappa-\omega}$. Using
	$\teff(L) \propto L^{-1/\mu}$ leads to
\begin{equation} \elabel{mu_case1}
\mu = \beta/(\omega-\kappa) >\nu \ .
\end{equation}
        This implies that $\teff(L)$ finally leaves the \FSS
	region, whose width scales like $L^{-1/\nu}$, toward the LTP.
\item[2)] $\kappa-\omega=-\beta/\nu$ (``correct''): In
	this case $Y'(x)$ remains constant, so that its argument
	either remains constant or vanishes, according to
	\eref{scaling_FSS}. Thus $\teff(L)$ decays at least as fast as
	$L^{-1/\nu}$, \ie $\mu\le\nu$. To the order considered here
	the equality applies.
\item[3)] $\kappa-\omega<-\beta/\nu$ (``too fast''): $Y'(x)$
	vanishes, following \eref{scaling_HTP}. For
	$Y'(k_{t}\teff(L)L^{1/\nu})\propto
	L^{\kappa-\omega+\beta/\nu}$ and $\teff \propto L^{-1/\mu}$
	this implies
\begin{equation}\elabel{mu_case3}
\mu = \frac{\gamma \nu}{2 \nu (\kappa-\omega + \beta/\nu) + \gamma} > \nu \ , 
\end{equation}
	provided that the
	denominator of \Eref{mu_case3} is positive.  Hence the model
	leaves the \FSS region toward the HTP.  The special case of
	negative $\mu$, implying divergent effective temperature, will
	be ignored.
\end{itemize}
Crucially, only for $\kappa-\omega=-\beta/\nu$ (case 2)) does the
model remain in the \FSS region.  To achieve this, $h/\epsilon$ must
be tuned \emph{exactly} in the way the order parameter scales in a
system displaying standard FSS, $\ave{|m|} \propto
L^{-\beta/\nu}$ while fixed at the critical temperature.  In all other 
cases the scaling of the effective
temperature eventually drives the model out of the \FSS region:
$\xi/L$ vanishes in the thermodynamic limit. Nevertheless, $\ave{T}$
converges to $T_c$, so that the correlation length
\begin{equation}
\xi \propto L^{\nu/\mu} 
\elabel{xil}
\end{equation}
diverges. With this scaling of $\xi$ 
\emph{all observables will show standard finite-size scaling with
$\nu$ replaced by $\mu$} \footnote{To be be precise, $\xi$ enters the
standard \FSS equations as $L^{\nu/\mu}$, which leads for example to
$\chi\propto L^{\gamma/\mu}$, $c_v\propto L^{\alpha/\mu}$, but also to
$\ave{|m|}\propto L^{\kappa-\omega}$. However $\kappa-\omega\ne
-\beta/\mu$ in the third case discussed above.}.

To illustrate the above analysis we performed simulations of an Ising
model with dynamics as described: 
Using Metropolis updating, the absolute magnetization
density is calculated after each scan over the lattice. According to
\eref{isingmotion} a new temperature is then calculated to be used in
the next sweep, $\dot{T} = -h + \ave{|m|} \epsilon$.
Starting from $T=2.27$, systems of size $L = 40, 80 ... 640$ were
updated at least $10^6$ times as transient and at least another $10^6$ times for
statistics.

Our numerical simulations fully confirm the above analysis: We observe
the standard \FSS exponents with $\nu$ replaced by $\mu$ for any
reasonable choice of $\kappa-\omega$.  The new scaling exponent $\mu$
(and $T_c$) can be determined from $\ave{T}-T_c \propto
L^{-1/\mu}$. Using it in an \FSS analysis allows us to identify all
standard critical exponents.  Even without the knowledge of $\mu$
three measurements, say $\alpha/\mu$, $\beta/\mu$ and $\gamma/\mu$,
are sufficient to determine all exponents using standard scaling laws.
This seems to defy common sense, since the exponents are determined
without referring to $T_c$, making it seemingly very attractive for
investigations of continuous phase transitions.  However, standard
methods, such as \FSS or analysis of the critical behavior at $\xi\ll
L$, are much more reliable and efficient: Not only is the above
identification of $\teff$ with the reduced temperature
questionable. More importantly, almost any choice of the scaling of
$h$ and $\epsilon$ leads to vanishing $\xi/L$. One therefore simulates
effectively independent patches of a lattice in a way that \FSS
effects remain important.

For two reasons the method is very sensitive to the choice of
$\epsilon_0$ and $h_0$ in \Eref{ratescaling} \footnote{As the dynamics
do not provide a natural timescale, a rescaling of these quantities
corresponds to a rescaling of time. In the limit of very slow dynamics
the situation of a fixed-temperature simulation is recovered.}:
Firstly, the amplitudes of the fluctuations in the effective
temperature depend directly on $h$ and $\epsilon$; choosing $h_0$ and
$\epsilon_0$ too large, the system destabilizes.
One can estimate these fluctuations by analyzing \eref{isingmotion}
and derive a lower bound for $\kappa,\omega$.  Secondly, if $h$ and
$\epsilon$ are too small and initially place the
system close to $T_c$, the scaling function reaches its asymptotic
behavior (generally \eref{scaling_HTP} or \eref{scaling_LTP}) only for
very large system sizes.

\begin{figure}
\includegraphics*[width=8.88cm,angle=0]{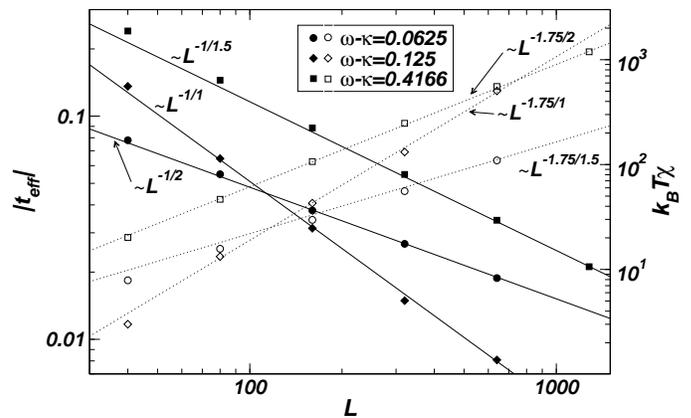}
\caption{The scaling of the effective temperature $\teff$ (filled symbols,
full lines) and the
susceptibility $\chi$ (open symbols, dashed lines) in the Ising model for
different choices of the exponent $\kappa$ and $\omega=1$. The symbols are
numerical simulations, the lines show the slopes expected from theory.}
\flabel{teff}
\end{figure}

\Fref{teff} shows the scaling of the effective temperature for the
three qualitatively different choices of the driving exponents
discussed above. The values of $\mu$ and $\gamma$ derived from these
data confirm the calculations. Depending on the choice of
$\kappa-\omega$, the value of $\mu$ immediately determines either
$\beta$, \Eref{mu_case1}, or $\gamma$, \Eref{mu_case3}. The \FSS of
specific heat and susceptibility produces the expected values of
$\alpha/\mu$ and $\gamma/\mu$.

Since our interest in the AS approach is due to its proposed role as
an explanation for SOC, we repeated the analysis for a variant of the
one-dimensional Abelian Manna model \cite{Manna:1991}.  This
sandpile-like model has been used to exemplify the link between SOC
and AS
\cite{DickmanVespignaniZapperi:1998,DickmanETAL:2000,VespignaniETAL:2000}.
It is driven in the bulk and implements bulk dissipation
\cite{BarratVespignaniZapperi:1999}, as suggested by \eref{motion}.
While the key equation \eref{freeenergy} has been confirmed
\cite{DickmanETAL:2001}, the Manna model is neither as well understood
nor as well-behaved as the Ising model. 
Unlike in the Ising model, the Manna model can get stuck when hitting an
inactive state. This complicates the analysis especially for the third case
discussed above. Nonetheless the
effective particle density $\ave{\zeta}$ is certainly a function of
$h$ and $\epsilon$ and therefore depends on their scaling with all the
consequences laid out above. Indeed, the numerics agrees fairly well
with the theoretical predictions. Most crucially, the activity as well
as the scaling of the avalanche size distribution show a clear,
immediate dependence on the choice of the two exponents $\omega$ and
$\kappa$.

One should note that introducing bulk-drive and -dissipation to SOC
models does not result in a full correspondence to AS models. Firstly,
there are important observables in SOC lacking a counterpart in AS and
vice versa.  For example, there is no obvious definition of the
avalanche size in the active phase of an AS model.  Similarly, the
definition of the SOC-activity is somewhat arbitrary in Abelian models
\cite{Dhar:1999}, due to the lack
of a unique microscopic timescale, which is needed when taking
temporal averages or measuring rates.  Secondly, for AS models
\Eref{freeenergy} contains the asymptotic conditional activity, while
in bulk-driven SOC models, such as the Manna model presented above,
the \emph{instantaneous} activity enters into the equation of motion
\eref{motion}.

The present study shows that the proposed explanation of SOC as
``self-organized'' AS criticality
\cite{DickmanVespignaniZapperi:1998,DickmanETAL:2000,Dickman:2002}
implies non-universal scaling behavior dependent on
$\kappa-\omega$. Universal, that is dissipation- and
driving-independent, scaling behavior cannot be achieved with the AS
approach. The question whether SOC models have universal features is
very important. Universality is a main justification for studying
simple models and for disregarding the details of the physical
processes they describe.

Despite the importance of this issue, it is still unclear whether SOC
systems can be grouped into universality classes; in fact, exponents
can change due to small changes in the update rules
(e.g. \cite{Malthe-Sorenssen:1999,OlamiFederChristensen:1992}), and
SOC is notorious for its wide variety of critical exponents.
Accepting the AS approach this would be a consequence of implicitly
setting the scaling of external drive and dissipation by the dynamical
rules of the different models. 

However, there is also strong evidence in favor of universality in
SOC. Many changes of the detailed dynamics do not affect the critical behavior
\cite{JensenPruessner:2003,BengrineETAL:1999,BengrineETAL:1999b,Zhang:1997}.
Moreover, the ratio $\xi/L$ appears to remain
constant in direct measurements of some models \cite{Peters:2004} so
that the ``correct'' \FSS exponents are observed, which is in stark
contrast to \eref{xil}.

At first sight, the observation of the same exponents in SOC and AS
models (such as
\cite{ChessaMarinariVespignani:1998,VespignaniETAL:2000,
ChristensenETAL:2004}) seems to support the case for the AS approach.
But our analysis shows that the opposite is true: If the AS approach
was determining the behaviour of SOC models, it would almost certainly
(apart from case 2)) produce exponents which \emph{differ} from those
observed in their AS counterparts.
Barring coincidence, observations supporting universality in SOC must
therefore be taken as strong evidence against the AS approach explaining 
the critical behavior of SOC models.

We have calculated the \FSS behaviour of a system approaching its
critical point through a feedback mechanism between order parameter
and tuning parameter.  While scale-free distributions of responses
such as those observed in the case of rainfall
\cite{PetersHertleinChristensen:2002} or earthquakes
\cite{GutenbergRichter:1944}, can be produced by such a process, it
only yields critical behavior strongly dependent on the detailed
dynamical rules of SOC models. There would be no universality and
robustness against small changes in the dynamical rules. 
While the AS mechanism in its present form may produce further insight
into potentially non-universal critical phenomena as observed
in field experiments, it fails to explain the apparent universality of
SOC models.

\begin{acknowledgments}
The authors gratefully acknowledge the support by EPSRC and would
like to thank Stefano Zapperi and Sven L\"{u}beck for helpful comments
on the manuscript.  GP would like to thank Alessandro Vespignani for
very useful discussions during NESPHY03 (MPIPKS Dresden), and the
the Alexander von Humboldt foundation as well as the NSF
(DMR-0088451/0414122) for support.
\end{acknowledgments}

\bibliography{ising_soc}
\end{document}